\documentclass[twocolumn,floatfix]{revtex4}

\usepackage{subfigure}
\usepackage{graphicx}

\begin{document}

\title{Patterns of Individual Shopping Behavior}

\author{Coco Krumme}
\author{Manuel Cebrian}
\author{Alex Pentland}
\affiliation{%
The Media Laboratory, Massachusetts Institute of Technology, Cambridge, MA 02139}%

\begin{abstract}
Much of economic theory is built on observations of aggregate, rather than individual, behavior. Here, we present novel findings on human shopping patterns at the resolution of a single purchase. Our results suggest that much of our seemingly elective activity is actually driven by simple routines. While the interleaving of shopping events creates randomness at the small scale, on the whole consumer behavior is largely predictable. We also examine income-dependent differences in how people shop, and find that wealthy individuals are more likely to bundle shopping trips. These results validate previous work on mobility from cell phone data, while describing the unpredictability of behavior at higher resolution. 

\end{abstract}

\maketitle

Human economic behavior is curbed by human geography: constraints on mobility determine where we can go and what we can buy. At the same time, the electivity of shopping itself drives our movement. Results from mobile phone data have shown that human mobility follows a truncated Levy flight \cite{gonzalez2008understanding}  and that the unpredictability of individual trajectories is bounded \cite{song2010limits}. But does this predictability hold from one dataset to the next? To what extent do existing routines drive where one shops, and to what extent do explicit decisions shape our shopping patterns? 

We consider a random sample of $10,000$ accounts from a major financial institution, over a three-month window in 2010. The anonymized dataset includes all check, credit card, and debit card transactions, as well as cash withdrawals and automatic wire payments. Merchants are identified by name and by Merchant Category Code (MCC) \footnote{Merchant Category Codes, available at {\tt http://www.irs.\\gov/irb/2004-31\_IRB/ar17.html}}, and we estimate the income of individuals from payments made into the account.  

\begin{figure}[ht]
\centering
\includegraphics[width=0.5\textwidth]{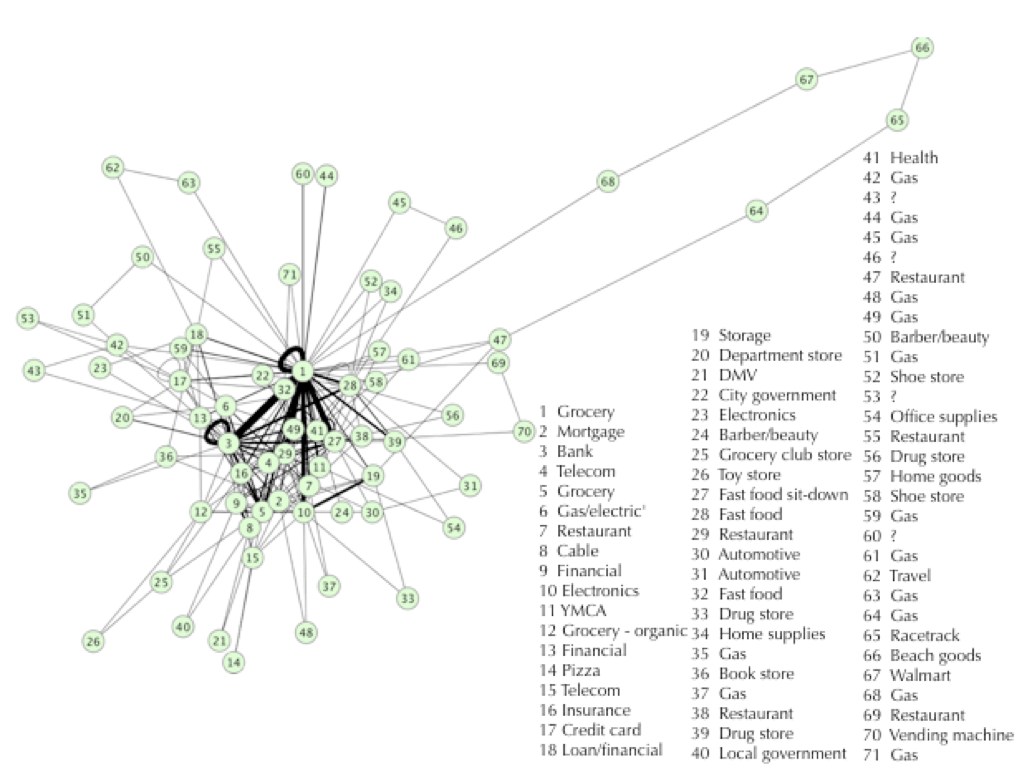}
\caption{Network linking stores frequented by a single individual over a three-months in 2008. Edge weight represents the transition probability from one location to another. A grocery store serves as the shopper’s main hub, and a natural grocery store as his secondary hub. Actual store names have been replaced with merchant type.}
\label{networkshops}
\end{figure}

\begin{figure}[ht]
\centering
\includegraphics[width=0.45\textwidth]{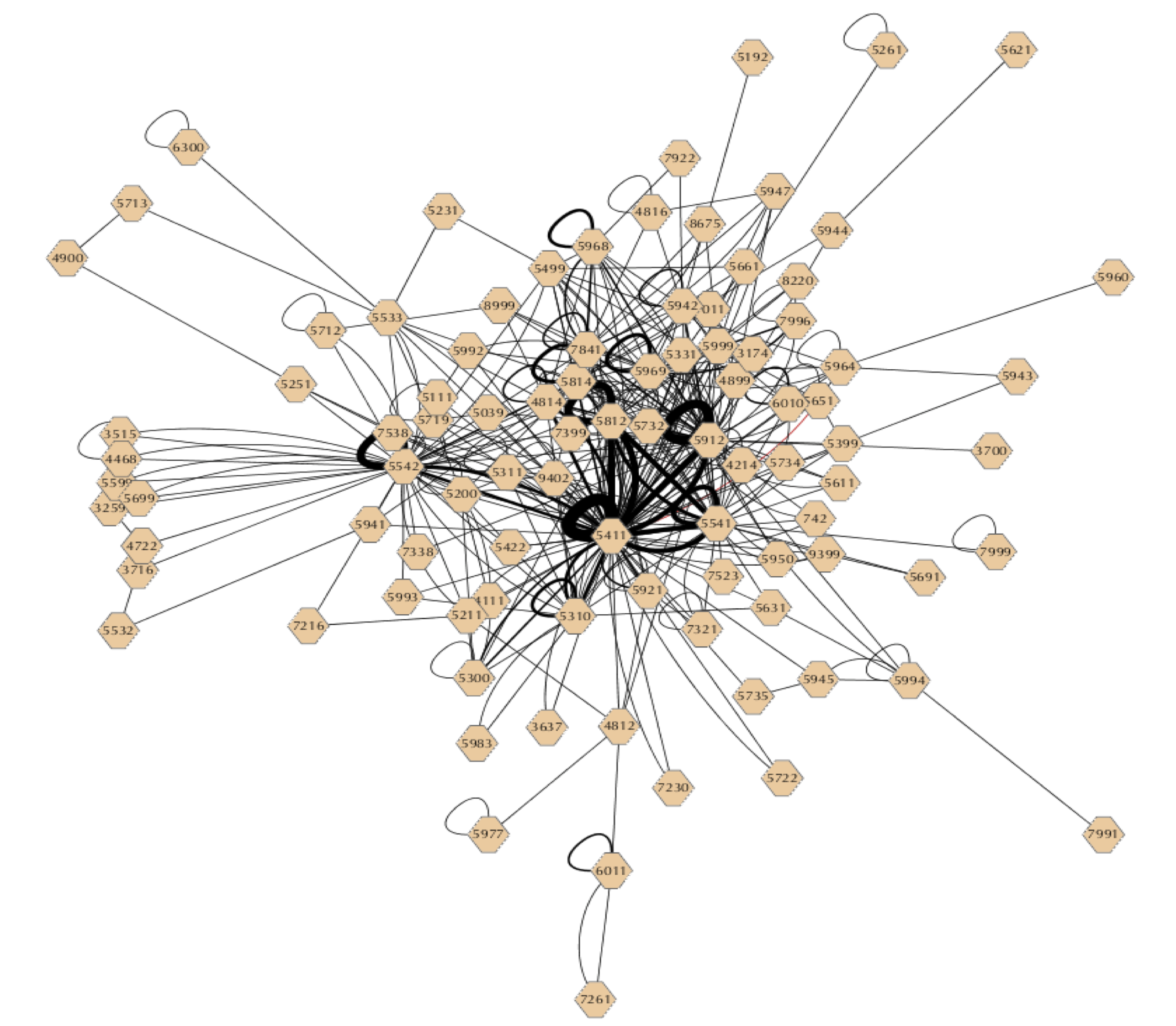}
\caption{Network of transitions between stores of different merchant category codes of $2,000$ individuals in 2008. Individuals belong to the top quintile of ``most predictable'' (lowest entropy, with total store visits held constant) in the sample. Edge weight and color represent the transition probabilities from a merchant in one retail category to one in another. For instance, {\tt 5411} represents grocery stores; a full list of MCC codes is available.}
\label{transitions}
\end{figure}

The sequence of shops frequented by a single individual can be described as a network where nodes are merchants and edge weights represent transition probabilities (fig. \ref{networkshops}). Shopping hubs represent locations of frequent purchase,  and are often grocery stores, gas stations or convenience stores. We are also able to plot population-level transition probabilities between merchants of different types (fig. \ref{transitions}).

\begin{figure}[ht]
\centering
\includegraphics[width=0.5\textwidth]{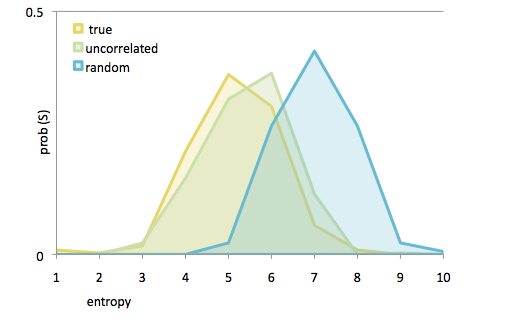}
\caption{Distribution of true, random and uncorrelated entropies for set of $2,000$ individuals over three-month window in 2010.  Relative to Song {\it et al.} results from mobile phone, adding sequence information does not lead to markedly lower distribution of true entropies.}
\label{distentropies}
\end{figure}

\begin{figure}[ht]
\centering
\includegraphics[width=0.45\textwidth]{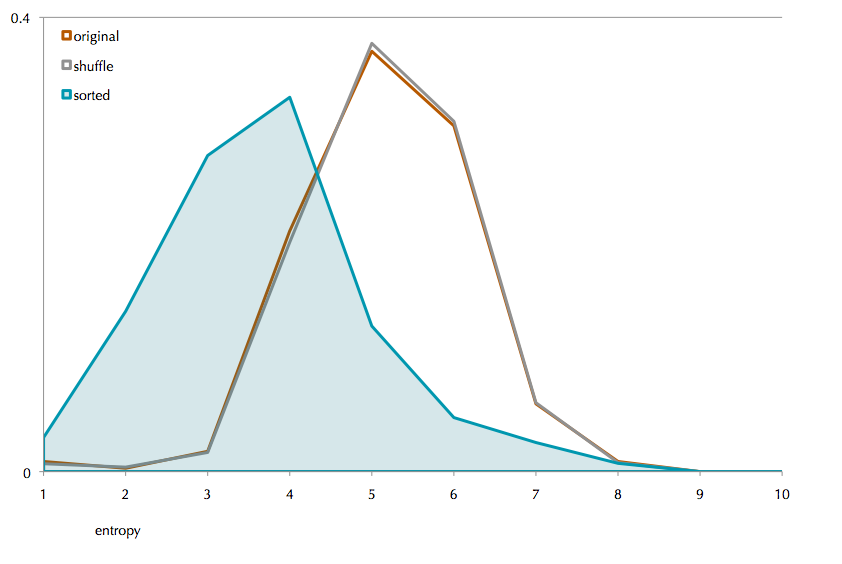}
\caption{$10,000$ run Monte Carlo simulation on a $2,000$ person subsample of effect of (a) randomizing and (b) sorting a set of stores visited over a week-long period. Averaging multiple runs of shuffled sequence information does not change the entropy significantly, while sorting lowers the entropy.}
\label{simulation}
\end{figure}

\begin{figure*}[!ht]
\centering
\subfigure[]{
\includegraphics[width=0.465\textwidth]{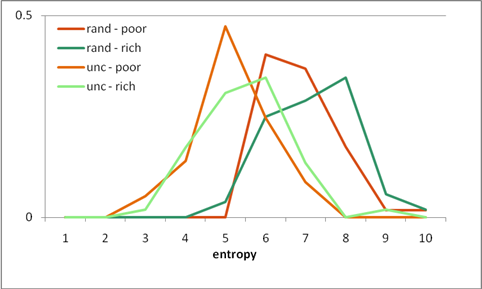}
\label{access}
}
\subfigure[]{
\includegraphics[width=0.45\textwidth]{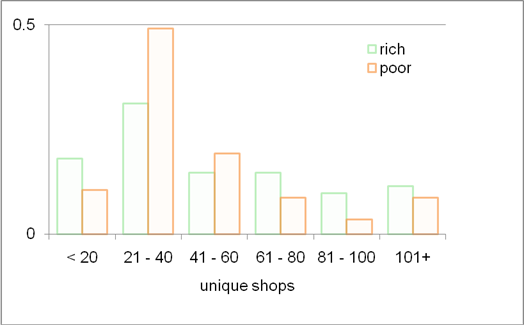}
\label{variety}
}
\newline
\subfigure[]{
\includegraphics[width=0.27\textwidth]{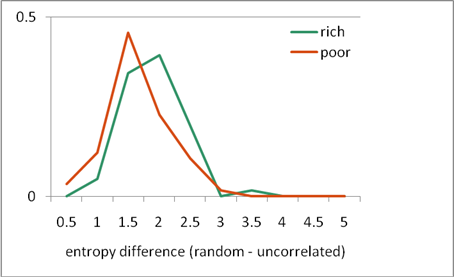}
\label{difference}
}
\subfigure[]{
\includegraphics[width=0.27\textwidth]{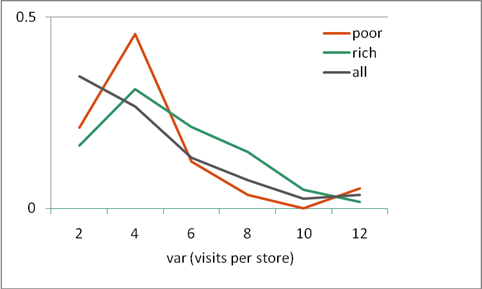}
\label{timestore}
}
\subfigure[]{
\includegraphics[width=0.27\textwidth]{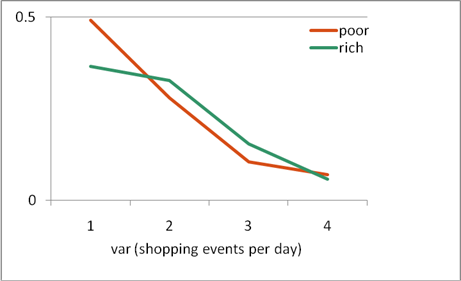}
\label{bundling}
}
\caption{(a) Distribution of number of stores frequented by sub-samples of wealthy and poor individuals; (b) random and temporally-uncorrelated entropies of sub-samples: the rich have higher overall entropies; (c) yet the by-person difference between random and uncorrelated entropies is greater for wealthy individuals, suggesting that variety in consumption patterns, rather than allotment over time, preferentially contributes to entropy for this group; (d) the greater variance in visits per store supports the notion that the entropy of the rich is due to preference for variety; (e) wealthy consumers are more likely to bundle together shopping events }
\label{decomposition}
\end{figure*}

Our findings suggest that at longer time scales, shopping behavior is constrained by some of the same features that have been seen to govern mobility generally. Shoppers return to stores with remarkable regularity: a Zipf's law ($s=4 \pm 0.031$) also describes the probability that a customer will visit a store at rank $N$ (where $N = 3$ is his third most-frequented store, for example), independent of the total number of stores visited in a three-month period. These results support those of Gonz\'alez {\it et al.}

Similarly, the random ($\log_{2} N_i$ , where $N_i$  = number of locations visited by shopper $i$) and temporal-uncorrelated ($\sum_i p_i (j) \log_2 p_i (j)$, where $p_i (j)$ is the probability that user $i$ visited location $j$) entropies of individuals comprise a distribution akin to that seen from cell phone patterns, albeit with higher mean entropies (fig. \ref{distentropies}). Moreover, these entropies are largely stable over four years for the same sample of individuals.

Yet this predictability falls off at higher temporal resolutions, suggesting that individual choices propel daily activity. When we incorporate the sequence of stores frequented to estimate the true entropy using the average Kolmogorov complexity, we find the entropy largely unchanged (fig. \ref{distentropies}): that is, the diversity of an individual's shopping behavior is driven by the number of stores he visits and the frequency at which he visits, not by the order in which he shops. This is a departure from the findings of Song {\it et al.}  in cell phone calls, where the order of sites visited explains a large portion of mobility patterns.

This discrepancy can be explained in part by the effect of {\it interleaving} store visits in time: I might go to the supermarket and then the post office, but I could just as well reverse this order. With the large-scale mobility patterns inferred from cell phones, I am unable to change many of my routines: I drive to the office after I drop off the kids at school. Indeed, when we simulate the effect of novel orderings by randomizing shopping sequence within a day, we find little change in total entropy. However, we are able to approach the levels of true entropy seen in the mobile phone data by sorting the order of shops visited over weekly intervals and thus imposing artificial regularity on shopping sequence (fig. \ref{simulation}).

We wish to understand, as well, whether there exist demographic correlates to these consumer behaviors. In contrast to literature on cell phones, which asserts that predictability is independent of features such as income, we observe marked differences in the ways that wealthy and poor individuals shop. We segment our sample and consider {\it poor} those with an annual inflow of less than \$$16,000$, and {\it wealthy} as those with inflow greater than \$$80,000$. Because we are only able to measure inflows into accounts associated with this bank, {\it income} here serves as a reasonable lower bound on earnings: other research suggests that this bound is asymmetric insofar as wealthy individuals are more likely to automatically direct income to multiple accounts \footnote{Federal Reserve Board 2007 Survey of Consumer Finances, available at {\tt http://www.federalreserve.gov/pubs/oss/\\oss2/2007/scf2007home.htm}}.

Wealthy people, on average, shop at a greater variety of stores than do their peers (fig. \ref{variety}), a phenomenon that might be attributed to differential access to commercial opportunities \cite{powell2007food}, to a taste for variety capacitated by greater discretionary income \cite{wood2005discretionary}, or to some combination of factors.  Because of the variety of their shopping, the rich have higher random entropies (fig. \ref{access}). An individual's temporally uncorrelated entropy, meanwhile, depends on the diversity of his store portfolio as well as on his distribution of purchases across stores.  

While the wealthy exhibit larger uncorrelated entropies than do the poor (fig. \ref{access}), the difference between random and uncorrelated entropies, on a by-person basis, is greater for the rich (fig. \ref{difference}). That is, a steeper distribution of store visits dampens the uncorrelated entropy of wealthy shoppers, while their tendency toward commercial variety increases entropy. Conversely, the regularity of purchases across shops is more important in explaining the predictability of the poor. These results are supported by the differential in variance of (times a store visited) between the rich and the rest (fig. \ref{timestore}).
 
Wealthy individuals are more likely to group purchases in time as well as in space. We propose the {\it bundling} of shopping trips as a behavioral explanation for this variance. We can think of bundling as a trip comprising visits to several stores, germane to the bursty behavior seen in other human systems \cite{barabasi2005origin}. Field studies \cite{lloyd1978shopping} have found that when choosing a grocery store, wealthy individuals optimize for proximity to other merchants (and thus make multiple purchases in the same trip), while poor individuals chose stores close to their homes.  

We define bundling as variance over bins representing daily activity, and observe a difference in distribution of bundling level between rich and poor individuals (fig.  \ref{bundling}). 

Holding income constant, we also find differences in customers with high and low entropies: while the most predictable individuals (entropy in highest quintile) tend to have grocery stores and restaurants as their topped-rank merchant, their unpredictable counterparts (entropy in last quintile) frequent fast food chains and gas stations. In simulation, there is a $31.4\%$ chance that two of the most or least predictable people will be found at the same merchant type for any given shopping event, but a predictable person will only overlap with an unpredictable person $11.2\%$ of the time.

Colloquially, an unpredictable person can exhibit one of several patterns: he may be hard to pin down, reliably late, or merely spontaneous. Yet given an individual's record, we seem to be able able to agree on whether he is predictable or not. As a more formal measure for human behavior, however, information-theoretic entropy conflates several of these notions. A person who discovers new shops and impulsively swipes his card presents a different case than the one who routinely distributes his purchases between his five favorite coffee shops, yet both have high entropy. Entropy remains a useful metric for comparisons between individuals and datasets (such as in the present study), but further work is need to tease out the correlates of predictability using measures aligned with observed behavior. 

In examining the solitary footprints that together comprise the invisible hand, we find that shopping, like mobility, is a highly predictable behavior at longer time scales. However, there exists substantial unpredictability in the sequence of shopping events over short time scales. While our results use an independent dataset to validate those of Song {\it et al.}, they also highlight the uniqueness of shopping behavior. A cell phone tower is a waypoint in an individual's daily trajectory, but a store is a destination, and ultimately, a nexus for human social and economic activity.

\bibliography{main}

\end{document}